\documentclass[aps,twocolumn,prd,superscriptaddress,showpacs,nofootinbib]{revtex4-1}

\usepackage[utf8]{inputenc}

\usepackage{diagbox}

\usepackage{mathtools}
\usepackage{amsfonts}
\usepackage{mathrsfs}
\usepackage{bbm}
\usepackage{slashed}

\usepackage{graphicx}
\usepackage{color}
\usepackage{array}

\usepackage{placeins}
\usepackage{booktabs}
\usepackage{makecell}
\usepackage{epstopdf}
\usepackage[caption=false]{subfig}

\usepackage{xspace}
\usepackage{siunitx}
\usepackage{hyperref}
\usepackage[nameinlink]{cleveref}
\usepackage{appendix}

\usepackage{xifthen}
\usepackage{xcolor}
\hypersetup{
	colorlinks,
	linkcolor={red!75!black},
	citecolor={blue!75!black},
	urlcolor={blue!75!black}
}

\begin{document}


\title{The rainbow modified-ladder approximation and degenerate pion}

\author{Lei Chang}
\email{leichang@nankai.edu.cn}
\affiliation{School of Physics, Nankai University, Tianjin 300071, China}

\author{Minghui Ding}
\email{mding@ectstar.eu} 
\affiliation{European Centre for Theoretical Studies in Nuclear Physics and Related Areas (ECT$^\ast$) and Fondazione Bruno Kessler\\ Villa Tambosi, Strada delle Tabarelle 286, I-38123 Villazzano (TN) Italy}


\begin{abstract}

Correlation functions can be described by the corresponding equations, $viz.$, gap equation for quark propagator and  the inhomogeneous Bethe-Salpeter equation for vector dressed- fermion-Abelian-gauge-boson vertex in which specific truncations have to be implemented. The general vector and axial-vector Ward-Green-Takahashi identities require these correlation functions to be interconnected, in consequence of this, truncations made must be controlled consistently. It turns out that if the rainbow approximation is assumed in gap equation, the scattering kernel in Bethe-Salpeter equation can adopt the ladder approximation, which is one of the most basic attempts to truncate the scattering kernel. Additionally, a modified-ladder approximation is also found to be a possible symmetry-preserving truncation scheme. As an illustration of this approximation for application a treatment of pion is included. Pion mass and decay constant are found to be degenerate in ladder and modified-ladder approximations, even though the Bethe-Salpeter amplitude are with apparent distinction. The justification for the modified-ladder approximation is examined with the help of the Gell-Mann-Oakes-Renner (GMOR) relation.

\end{abstract}

\maketitle

\section{\label{sec:section1}Introduction}

Hadron is a composite particle consist of quarks and gluons that are strongly interacting, so it cannot be described by perturbation theory. Rather, constituent quark model and parton model had been historically linked to the instructive description of hadron, so much so that it seems their development are progressively paving our way to gain greater insight directly from quantum chromodynamics (QCD) ~\cite{GellMann:1964nj,Zweig:352337,*Zweig:570209,Feynman:1969wa,Bjorken:1969ja}. Yet how does QCD give rise to the physics of hadron is a question that hasn't been mathematically answered. This intractable aspect of the theory is investigated by Lattice QCD, which has already successfully agreed with many experiments on non-perturbative phenomena~\cite{Aoki:2019cca}. Meanwhile, continuum field approach can serve as a complementary, such as Dyson-Schwinger equations (DSEs)~\cite{Roberts:1994dr,Maris:2003vk,Eichmann:2016yit} and the functional renormalisation group (fRG)~\cite{Pawlowski:2005xe,Cyrol:2017ewj}. More precisely, DSE, for instance, is a non-perturbative approach that can manifest main properties of QCD, dynamical chiral symmetry breaking (DCSB) and color confinement. It is yielding progress on hadron phenomena, in particular that of the lightest meson - pion. Despite being a bound state, pion is the pseudo Nambu-Goldstone mode generated by DCSB, and this dichotomous character entails that it takes a special position among theoretical interests on hadron~\cite{Maris:1997hd,Dai:1990ap}. The point to maintain this dichotomous character of pion is obeying the fundamental symmetry in QCD. That in DSE is to follow some basis rules when truncating the infinite coupled equations for correlation functions~\cite{Maskawa:1974vs,Maskawa:1975hx}. Herein the starting points are quark gap equation and the Bethe-Salpeter equations (BSEs).

The rainbow approximation for quark gap equation and the ladder approximation for BSE had been developed for decades ~\cite{Salpeter:1951sz,Mandelstam:1979xd}, and applied to a wide range of hadronic systems, including meson, baryon, and exotic state, etc.~\cite{Roberts:1994dr,Maris:2003vk,Eichmann:2016yit}. The underlying laws governing their application in a simple and elegant way is the preserving of vector and axial-vector Ward-Green-Takahashi identities driven by the gauge symmetry. They lead to peculiar relationships between truncations made for quark gluon vertex in gap equation and the scattering kernel in the Bethe-Salpeter equation. Two nonlinear equations, which connect the quark gluon vertex with the scattering kernel, can be acquired to exhibit the symmetry-driven correspondence. Starting from these two equations, one notice that if the quark gluon vertex is assumed to be a bare vertex, $i.e.$, the rainbow approximation as it is usually called, then a straightforward expression for the scattering kernel is immediately realised to be valid, which is interpreted as the ladder approximation. Given the existence of this nontrivial solution, $i.e.$, rainbow ladder approximation, for two nonlinear equations, one would be inclined to expect other solutions may also exist. 

Attempts can be made on two directions searching for other solutions. One of the directions is to go beyond both rainbow and ladder approximations, consistently truncating quark gluon vertex in DSE and scattering kernel in BSE ~\cite{Watson:2004kd,Matevosyan:2006bk,Fischer:2007ze,Fischer:2008wy,Heupel:2014ina,Qin:2016fbu,Qin:2016fwx,Binosi:2016rxz}. It turns out that this direction is indeed workable in practice. Additionally, it is proved that if the quark gluon vertex is added correction to the bare vertex, $viz.$, beyond rainbow approximation, the Nambu-Goldstone theorem is manifest only if scattering kernel is constructed consistently~\cite{Bender:1996bb}. With the extension of rainbow ladder approximation on this direction, one may acquire a good description of the spin-orbit splitting in the light meson~\cite{Chang:2009zb,Williams:2009wx},  the level ordering of pseudoscalar and vector meson radial excitations~\cite{Qin:2020jig}, and heavy-light meson mass spectrum~\cite{Qin:2019oar}. One may acquire further potential access charactering meson internal structures, such as distribution amplitude~\cite{Chang:2013pq} and resonance width~\cite{Williams:2018adr,Eichmann:2020oqt}.

An alternative direction is to keep rainbow approximation, instead modifying the ladder approximation~\cite{Chen:2019otg,Chang:2019eob}. We seek for this possibility in this work, assuming the ladder approximation can include a multiplicative factor, which can be recognised as our ${Ans\ddot{a}tze}$ for the scattering kernel. This introduced factor will be determined by the two nonlinear equations constrained from the preserving of vector and axial-vector Ward-Green-Takahashi identities. In this case, through imposing this nontrivial multiplicative factor, we find two other possible solutions for the two nonlinear equations. The procedure of expressing the scattering kernel in terms of ladder approximation with a multiplicative factor, and deriving its well-constrained form is a mathematical procedure, independent of the system under consideration. To explore whether it is useful in practical system, we include its application on pion, examining whether the Nambu-Goldstone theorem is manifest with this modified-ladder approximation, as well as its impact on the internal structure of pion. Properties such as mass, Bethe-Salpeter amplitude and decay constant are considered in comparison with those in rainbow ladder approximation. Additional examination of the reasonability for modified-ladder approximation is provided by the verification of the GMOR relation.

The remainder of this paper is organised as follows. In
Sec. \ref{sec:section2} we reiterate the vector and axial-vector Ward-Green-Takahashi identities, highlighting two corresponding nonlinear equations between quark gluon vertex and the scattering kernel. Sec. \ref{sec:section3} introduces the rainbow ladder approximation together with the rainbow modified-ladder approximation. Sec. \ref{sec:section4} deals with the application of rainbow modified-ladder approximation on pion.  Sec. \ref{sec:section5} contains our results of pion mass, Bethe-Salpeter amplitude, decay constant, as well as the discussion on the GMOR relation. Finally, we summarise in Sec. \ref{sec:section6}.

\section{\label{sec:section2}vector and axial-vector \protect\\Ward-Green-Takahashi identities}

Quantum chromodynamics (QCD) provides underlying laws governing the properties of particles, both elementary and hadronic. It requires the correlation functions to be interconnected by the Slavnov-Taylor identity, which corresponds to the Ward-Green-Takahashi identity in a Abelian gauge. We consider herein the connections between the $3$-point vertex of a dressed-fermion to an Abelian gauge boson, and $2$-point function, $i.e.$, quark propagator, and they are expressed by the vector and axial-vector Ward-Green-Takahashi identities. 

\subsection{\label{sec:section2sub1}Vector Ward-Green-Takahashi identity}
Ward identity~\cite{Ward:1950xp}, and its generalization by Green~\cite{Green:1953te} and Takahashi~\cite{Takahashi:1957xn}, is
\begin{equation}\label{eq:vwti}
iP_{\mu}\Gamma_{\mu}(k;P)=S^{-1}(k_+)-S^{-1}(k_{-})\,,
\end{equation}
which relates $3$-point function, the vector dressed-fermion-Abelian-gauge-boson vertex $\Gamma_{\mu}(k;P)$ with $2$-point function, the quark propagator $S(k)$. The original Ward identity, derived earlier by Ward from a study of perturbation theory, can be obtained from letting $k_+$ approach $k_-$, with $k_\pm=k\pm P/2$, and $k$, $P$ are respectively, relative and total momentum of the dressed quark and dressed antiquark.  

The quark propagator in Eq.\eqref{eq:vwti} satisfies the gap equation\footnote{We use a Euclidean metric: $\{\gamma_\mu,\gamma_\nu\} = 2\delta_{\mu\nu}$; $\gamma_\mu^\dagger = \gamma_\mu$; $\gamma_5= \gamma_4\gamma_1\gamma_2\gamma_3$, tr$[\gamma_4\gamma_\mu\gamma_\nu\gamma_\rho\gamma_\sigma]=-4 \epsilon_{\mu\nu\rho\sigma}$; $\sigma_{\mu\nu}=(i/2)[\gamma_\mu,\gamma_\nu]$; $a \cdot b = \sum_{i=1}^4 a_i b_i$; and $P_\mu$ timelike $\Rightarrow$ $P^2<0$.  More information is available in Sec.\,2.3 of Ref.\,\protect\cite{Roberts:1994dr}.} 
\begin{align}\label{eq:gap}
&S^{-1}(k)=Z_2 (i\gamma{\cdot}k+ Z_m m^\zeta)\notag\\
&+Z^2_2\int_{dq}g^2\mathcal{D}_{\mu\nu}(k-q)\frac{\lambda^a}{2}\gamma_{\mu}S(q)\frac{\lambda^a}{2}\Gamma_\nu(k,q)\,,
\end{align}
where $m^\zeta$ is the current quark mass, and $\zeta$ is the renormalisation scale; $Z_{2,m}$, respectively, quark wave function and mass renormalisation constants; $\mathcal{D}_{\mu\nu}(k-q)$ the gluon propagator; $\Gamma_\nu(k,q)$ the quark gluon vertex; $\lambda^a$ the hermitian Gell-Mann matrices.

As outlined in the preceding introduction, we seek for a possible modified-ladder approximation with keeping rainbow approximation. In the following discussion, we will implement the rainbow approximation in gap equation, and in this case, the quark gluon vertex in Eq.\eqref{eq:gap} is the bare vertex as
\begin{equation}\label{eq:rainbow}
	\Gamma_\nu(k,q)=\gamma_\nu\,.
\end{equation}

The $3$-point function with specified $J^P$ quantum number  satisfies the general inhomogeneous Bethe-Salpeter equation
\begin{align}\label{eq:vectorbse}
&[\Gamma_{J^P}(k;P)]_{\alpha\beta}
=Z_{J^{P}}[\gamma_{J^P}]_{\alpha\beta}\notag\\
+&Z^2_2\int_{dq}[\mathcal{K}(k,q,P)]_{\alpha\alpha^\prime;\beta^\prime\beta}[S(q_+)\Gamma_{J^P}(q,P)S(q_-)]_{\alpha^\prime\beta^\prime}\,,
\end{align}
with $\mathcal{K}(k,q,P)$ is the quark-antiquark scattering kernel. $\gamma_{J^P}={\bf{1}},\gamma_5,\gamma_\mu,\gamma_5\gamma_\mu$, and ect., is the inhomogeneous driving term corresponding to the $J^P$ quantum number, and $Z_{J^{P}}$ is the related renormalisation constant. The vertex appeared in Eq.\eqref{eq:vwti} is the vector dressed-fermion-Abelian-gauge-boson vertex with $\gamma_{J^P}=\gamma_\mu$, and $Z_{J^{P}}=Z_2$.  With the specification of quark propagator $S(k)$ and vector vertex $\Gamma_{\mu}(k;P)$ in Eq.\eqref{eq:vwti}, as well as the implication of the rainbow approximation, the only quantity that unknown, $\mathcal{K}(k,q,P)$ can be well settled.

Inserting the quark propagator in Eq.\eqref{eq:gap} and the vector vertex in Eq.\eqref{eq:vectorbse} into the vector Ward-Green-Takahashi identity expressing their connection in Eq.\eqref{eq:vwti}, one will have
\begin{eqnarray}\label{eq:relationfromvwti}
&&\int_{dq}\mathcal{K}(k,q,P)_{\alpha\alpha^\prime;\beta^\prime\beta}[S(q_+)-S(q_{-})]_{\alpha^\prime\beta^\prime}\notag\\
=&&-\int_{dq}g^2\mathcal{D}_{\mu\nu}(k-q)\frac{\lambda^a}{2}\gamma_{\mu}[S(q_{+})-S(q_{-})]\frac{\lambda^a}{2}\gamma_{\nu}\,.
\end{eqnarray}
This is a relation that scattering kernel $\mathcal{K}(k,q,P)$ must preserve in the rainbow approximation.

\subsection{\label{sec:section2sub2}Axial-vector Ward-Green-Takahashi identity}
The other generalised Ward identity is
\begin{align}\label{eq:avwti}
P_{\mu}\Gamma_{5\mu}(k;P)=&S^{-1}(k_+)i\gamma_5+i\gamma_5S^{-1}(k_{-})\notag\\
&-2im^\zeta\Gamma_5(k;P)\,,
\end{align}
where $\Gamma_{5\mu}(k;P)$ is the axial-vector vertex, and it  satisfies the inhomogeneous  axial-vector Bethe-Salpeter equation in Eq.\eqref{eq:vectorbse} with $\gamma_{J^P}=\gamma_5\gamma_\mu$ and $Z_{J^{P}}=Z_2$; $\Gamma_5(k;P)$ is the pseudoscalar vertex, and it satisfies the inhomogeneous pseudoscalar Bethe-Salpeter equation in Eq.\eqref{eq:vectorbse} with $\gamma_{J^P}=\gamma_5$ and $Z_{J^{P}}=Z_4$.

Inserting the quark propagator in Eq.\eqref{eq:gap},  axial-vector and pseudoscalar vertices in Eq.\eqref{eq:vectorbse} into the axial-vector Ward-Green-Takahashi identity in Eq.\eqref{eq:avwti}, one will have
\begin{eqnarray}\label{eq:relationfromavwti}
&&\int_{dq}\mathcal{K}(k,q,P)_{\alpha\alpha^\prime;\beta^\prime\beta}[S(q_+)\gamma_5+\gamma_5S(q_{-})]_{\alpha^\prime\beta^\prime}\notag\\
=&&-\int_{dq}g^2\mathcal{D}_{\mu\nu}(k-q)\frac{\lambda^a}{2}\gamma_{\mu}[S(q_{+})\gamma_5+\gamma_5S(q_{-})]\frac{\lambda^a}{2}\gamma_{\nu}\,.
\end{eqnarray}
This is a second relation that scattering kernel $\mathcal{K}(k,q,P)$ must preserve in the rainbow approximation.

In this way we see that, the use of vector and axial-vector Ward-Green-Takahashi identities in the construction of two general relations: Eq.\eqref{eq:relationfromvwti} and Eq.\eqref{eq:relationfromavwti} for the scattering kernel in the rainbow approximation has been motivated by the requirement of gauge symmetry and chiral symmetry. These two constructed relations in the case of rainbow approximation are consistent with those equations derived in a more general case~\cite{Qin:2016fwx,Qin:2016fbu,Qin:2020jig}. Notably, the procedure of constructing the scattering kernel is a mathematical procedure, independent of the system under consideration.

\section{\label{sec:section3}Quark-antiquark scattering kernel}

With the recognition of mathematical formulation associated with gauge and chiral symmetries in Eq.\eqref{eq:relationfromvwti} and Eq.\eqref{eq:relationfromavwti} must be preserved for the scattering kernel, we shall begin to solve  these two equations and the resulting scattering kernel will automatically  keep corresponding symmetries.

\subsection{\label{sec:section3sub1} Ladder approximation}

There is one apparent practical solution, which is 
\begin{eqnarray}\label{eq:rainbowladder}
	&&[\mathcal{K}(k,q,P)]^{\text{RL}}_{\alpha\alpha^\prime;\beta^\prime\beta}\notag\\
	=&&-g^2\mathcal{D}_{\mu\nu}(k-q)\left[\frac{\lambda^a}{2}\gamma_{\mu}\right]_{\alpha\alpha^\prime}\otimes\left[\frac{\lambda^a}{2}\gamma_{\nu}\right]_{\beta^\prime\beta}\,.
\end{eqnarray}
It is known as the rainbow ladder (RL) approximation. The ladder approximation has been one of the most successful attempts to truncate Dyson-Schwinger equations, which is still meaningful in a plenty of recent studies~\cite{Eichmann:2019tjk,Eichmann:2019bqf,Gutierrez-Guerrero:2019uwa,Frederico:2019noo,Ding:2019lwe,Ding:2019qlr,Gao:2020hwo,Barabanov:2020jvn,Aguilar:2019teb}. From the mathematical viewpoint, it is recognised as the leading order truncation scheme, and any correction to it can provide us further assistance in our better understanding of the truncations.

\subsection{\label{sec:section3sub2} Modified-ladder approximation}

 If one solution of a nonlinear system of equations exists, it is often possible to find several solutions and we realise this is the case herein. 
 We now assume that the scattering kernel in Eq.\eqref{eq:rainbowladder} has an extension as
 \begin{eqnarray}\label{eq:rainbowmodifiedladder}
&& [\mathcal{K}(k,q,P)]^{\text{RML}}_{\alpha\alpha^\prime;\beta^\prime\beta}\notag\\
&& =-g^2\mathcal{D}_{\mu\nu}(k-q)\left[\frac{\lambda^a}{2}\gamma_{\mu}\Lambda_\beta\right]_{\alpha\alpha^\prime}\otimes\left[\Lambda_\beta\frac{\lambda^a}{2}\gamma_{\nu}\right]_{\beta^\prime\beta}\,,
 \end{eqnarray}
 where $\Lambda_\beta$ is a Dirac structure function. Thus the scattering kernel is performed to be the ladder approximation with a multiplicative factor. We have to admit that this is an ${Ans\ddot{a}tze}$, and whether it is reliable remains to be justified by whether we can find a nontrivial expression for $\Lambda_\beta$. The resulting $\Lambda_\beta$ must be distinguished from $\bf{1}$, so that modified-ladder approximation is distinguished from ladder approximation.
 
 The procedure finding nontrivial $\Lambda_\beta$ is a mathematical procedure. In general, $\Lambda_\beta$ can be of all Dirac matrices
$\left\{\gamma_5,\frac{1}{2}\gamma_{\beta}\,,\frac{i\gamma_5}{2}\gamma_{\beta}\,,\frac{i}{2\sqrt{3}}\sigma_{\beta\alpha}\,, \frac{i\gamma_5}{2\sqrt{3}}\sigma_{\beta\alpha}\,\right\}$~\cite{arflcen1985mathematical}, however practical calculation yields only two possible following forms
\begin{eqnarray}\label{eq:models}
\Lambda^+_\beta=&&\frac{i}{\sqrt{3}}\frac{\gamma_5\sigma_{\beta\alpha}(iq_{+}^\alpha\sigma_v(q_{+}^2)-iq_{-}^\alpha\sigma_v(q_{-}^2))}{\sqrt{q_{+}^2\sigma^2_v(q_{+}^2)+q_{-}^2\sigma^2_v(q_{-}^2)-2q_{+}{\cdot}q_{-}\sigma_v(q_{+}^2)\sigma_v(q_{-}^2)}}\,,\nonumber\\
\Lambda^-=&&\frac{1}{2}\frac{\gamma_5\sigma_{\beta\alpha}(iq^\alpha_{+}q^\beta_--iq^\alpha_-q^\beta_+)}{\sqrt{(q_+{\cdot}q_-)^2-q^2_+q^2_-}}\,,
\end{eqnarray}
where $\sigma_{\mu\nu}=\frac{i}{2}(\gamma_{\mu}\gamma_{\nu}-\gamma_{\nu}\gamma_{\mu})$ and $q_{\pm}=q\pm P/2$; $\sigma_v$ is the vector part of quark propagator 
$S(q)=-i\gamma{\cdot}q\sigma_v(q^2)+\sigma_s(q^2)$.

The expressions in Eq.\eqref{eq:models} show up very clearly the difference between ladder approximation and modified-ladder approximation. In the case of ladder approximation for which there is no momentum dependence in scattering kernel $\mathcal{K}(k,q,P)$ expect for that appears in gluon propagator. However, there is momentum dependence in modified-ladder approximation, described by quark momentum $q_{\pm}$ and/or the vector part of quark propagator $\sigma_v(q^2)$. In the consequence of this, the modification on the scattering kernel to some extent can be considered as the rearrangement of the quark momentum within hadron.

Additionally, given the importance of the examination of whether the Nambu-Goldstone theorem is manifest in modified-ladder approximation, we specify the formulation of the expression $\Lambda^+_\beta$ in Eq.\eqref{eq:models} in chiral limit. If a massless pion exist, then $P=0$ and $\Lambda^+_\beta=\frac{i\gamma_5}{\sqrt{3}}\sigma_{\beta\alpha}n_\alpha$, with $n=(0,0,0,i)$ is the unit vector in the direction of $P$.

\begin{figure}[t]
\centerline{\includegraphics[width=0.2\textwidth]{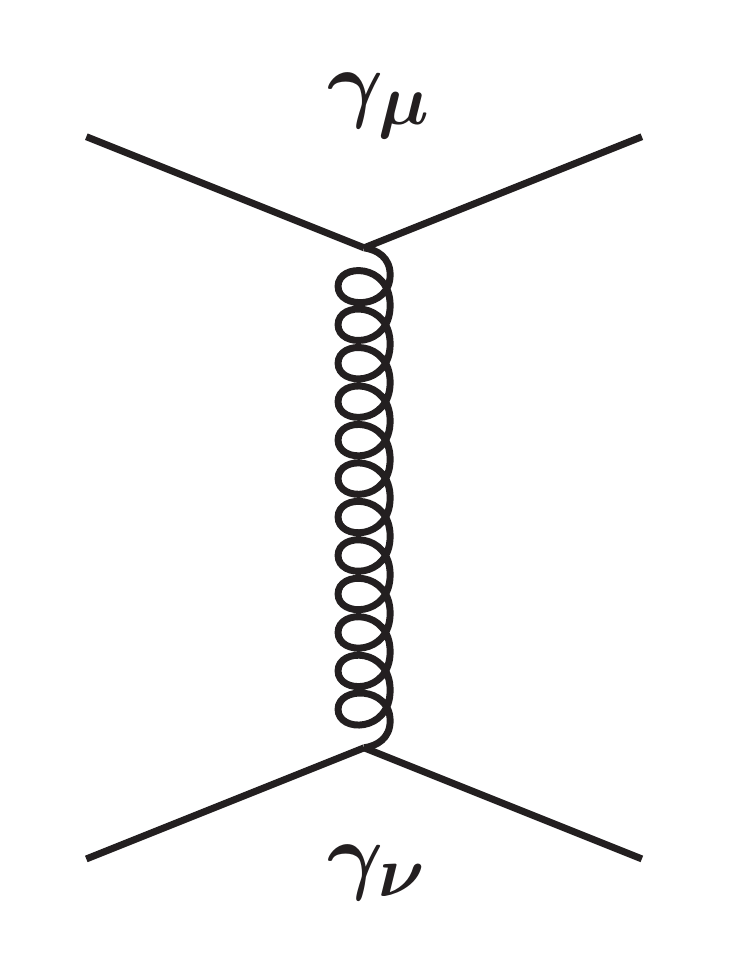}
\includegraphics[width=0.22\textwidth]{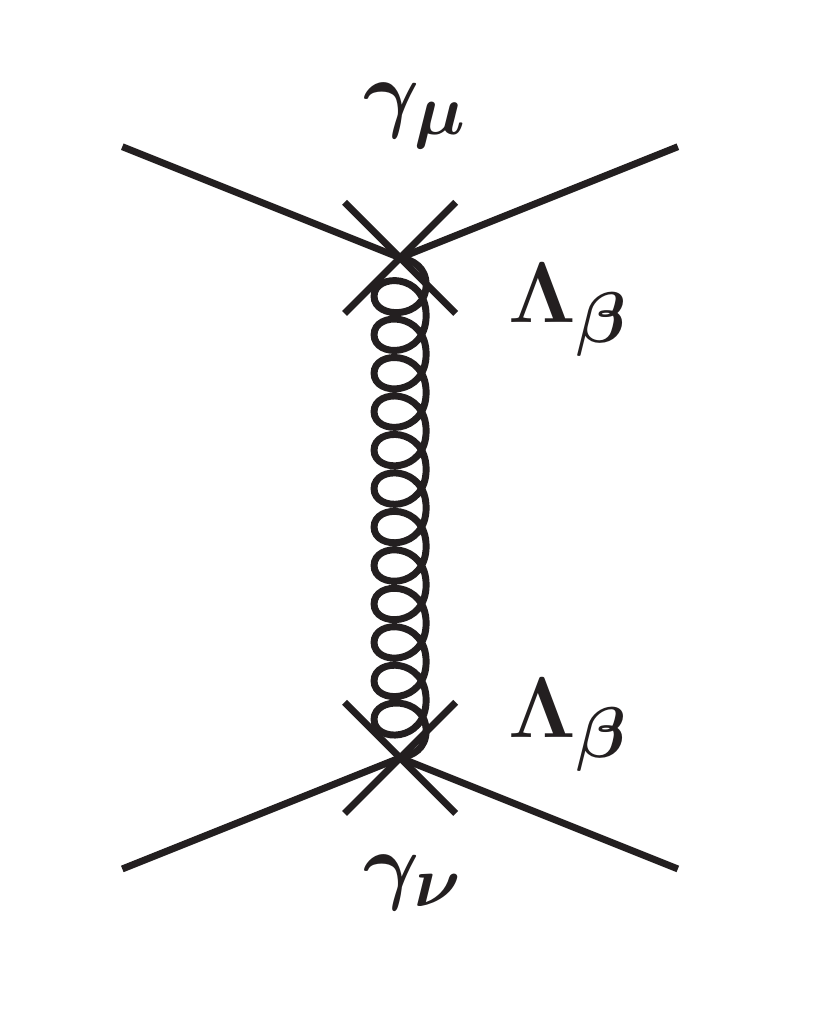}}
\caption{$\textit{Left panel}$ - Scattering kernel with ladder approximation in Eq.\eqref{eq:rainbowladder}. $\textit{Right panel}$ - Scattering kernel with modified-ladder approximation in Eq.\eqref{eq:rainbowmodifiedladder}. The internal solid lines represent dressed gluon propagators. The cross indicates to multiply a $\Lambda_\beta$ on where it is marked.} \label{fig:kernelfeymann}
\end{figure}

\section{\label{sec:section4}Application: pion}

In the preceding section, we have found two mathematical solutions for rainbow modified-ladder (RML) approximation, and in order to justify their rationality, we shall now consider their application on the lightest hadron system, pion. The study on pion is an apparent direction, since it is possible to consider whether the Nambu-Goldstone theorem is manifest with RML approximations, which is the basic criteria that one must meet when developing new scattering kernels~\cite{Bender:1996bb}. The analysis on pion mass, decay constant and Bethe-Salpeter amplitude should be quite interesting affairs. It is known that pion can be described by the homogeneous Bethe-Salpeter equation
\begin{eqnarray}\label{eq:pionbse}
	&&\lambda(P^2)S^{-1}(k_+)[\chi(k;P)]_{\alpha\beta}S^{-1}(k_-)\notag\\
		= &&Z^2_2 \int_{dq} [\mathcal{K}(k,q,P)]_{\alpha\alpha^\prime;\beta^\prime\beta} [\chi(q,P)]_{\alpha^\prime\beta^\prime}\,,
\end{eqnarray}
with $\lambda(P^2)$ is the eigenvalue of the kernel, which is equal to one when pion is on-shell, $\lambda(P^2=-M_{\pi_i}^2)=1$. The corresponding eigenvector of Eq.\eqref{eq:pionbse} is pion Bethe-Salpeter wave function $\chi(k,P)$. When solving this equation, we take pion Bethe-Salpeter wave function as
\begin{eqnarray}	
\chi(k;P)=\sum_{i=1}^4 \tau_i(k;P) f_i(k;P)\,,
\end{eqnarray}
where $f_i(k,P)$ is scalar function characterising pion internal wave function dependence on relative and total momentum of the dressed quark and dressed antiquark. $\tau_i$ is the complete set of the Dirac bases for pseudoscalar meson Bethe-Salpeter wave function~\cite{Maris:1997hd}
\begin{eqnarray}\label{eq:pionbsebasis}
&&\tau_1=i\gamma_5\,,\quad\quad\quad\quad\quad\,\,\,\,\tau_2=\gamma_5\gamma{\cdot}P\,,\notag\\
&&\tau_3=\gamma_5P{\cdot}k\gamma{\cdot}k\,,\quad\quad\quad\tau_4=\gamma_5\sigma_{\mu\nu}k_\mu P_\nu\,.
\end{eqnarray}

By adapting this set of Dirac bases, we find special properties with RML approximations. For instance, if considering $\tau_i$ multiply the Lorentz structure of the right hand side of Eq.\eqref{eq:pionbse} in the case of $\Lambda^+_\beta$, which is the common procedure when solving Eq.\eqref{eq:pionbse}, one may immediately notice  
\begin{eqnarray}\label{lambdaplustrace}
\text{tr}[\tau_i\Lambda^+_\beta\tau_j\Lambda^+_\beta]&&=\text{tr}[\tau_i\tau_j]\,,\quad|i=1\nonumber\\
\text{tr}[\tau_i\Lambda^+_\beta\tau_j\Lambda^+_\beta]&&\neq\text{tr}[\tau_i\tau_j]\,,\quad|i=2,3,4
\end{eqnarray}
with $j$ being any number. This indicates that equations for $f_1(k;P)$ with RML in the case of $\Lambda^+_\beta$ are equivalent to those with RL, whereas equations for $f_{2,3,4}(k;P)$ are different from RL ones. Further, if considering $\tau_i$ multiply the Lorentz structure of the right hand side of Eq.\eqref{eq:pionbse} in the case of $\Lambda^-$, one can get 
\begin{eqnarray}
\text{tr}[\tau_i\Lambda^-\tau_j\Lambda^-]&&=\text{tr}[\tau_i\tau_j]\,,\quad|i=1\nonumber\\
	\Lambda^-\tau_i\Lambda^-&&=\tau_i\,,\quad\quad\quad\,|i=2,3,4
\end{eqnarray}
for any $j$. Thus all equations in the case of $\Lambda^-$ resemble those with RL. Therefore, we can even imagine before calculation that we will obtain eigenvalues and eigenvectors of the Bethe-Salpeter kernel in the case of $\Lambda^-$ the same as their corresponding RL ones. This property makes us conclude that RML in the case of $\Lambda^-$ for pion is equivalent to RL. For this reason, the focus of this paper is on RML in the case of $\Lambda^+_\beta$.
 
The procedure solving Eq.\eqref{eq:pionbse} is the common process finding eigenvalues and eigenvectors of the Bethe-Salpeter kernel~\cite{Sanchis-Alepuz:2017jjd}. After locating pion mass, we simultaneously obtain pion on-shell Bethe-Salpeter wave function, and it is the eigenvector associated with eigenvalue $\lambda(P^2=-M_{\pi_i}^2)=1$. The Bethe-Salpeter wave function can be normalised by the condition
\begin{eqnarray}\label{eq:norm}
	\left(\frac{\partial\ln(\lambda)}{\partial{P}^2}\right)^{-1}=\text{tr}\int_{dq}\bar{\Gamma}(q;-P)\chi(q;P)\,,
\end{eqnarray}
with $\lambda$ is the eigenvalue; $\bar{\Gamma}(q;P)=\hat{C}\Gamma^t(-q;P)\hat{C}^{-1}$ is the charge conjugation of Bethe-Salpeter amplitude $\Gamma(q;P)=S^{-1}(q_+)\chi(q;P)S^{-1}(q_-)$~\cite{Nakanishi:1965zza,Nakanishi:1965zz}.
  
\begin{figure}[t]
\centerline{\includegraphics[width=0.5\textwidth]{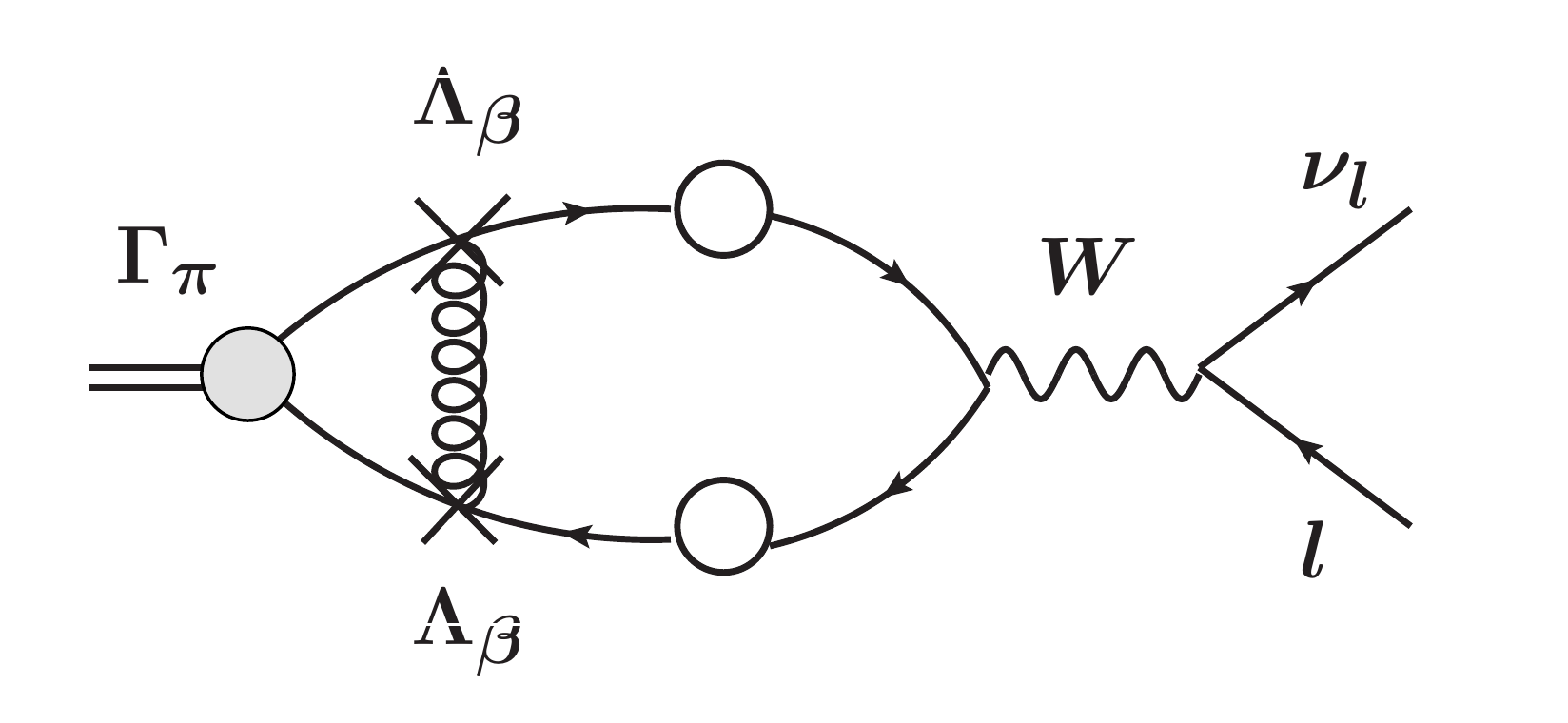}}
\caption{Pion decay constant with modified-ladder approximation in Eq.\eqref{eq:piondecay}. Filled circle represents pion Bethe-Salpeter amplitude, and hollow circles are dressed quark propagators. The cross indicates to multiply a $\Lambda_\beta$ on where it is marked.} \label{fig:decayfeymann}
\end{figure}

The normalised Bethe-Salpeter wave function can be used to define pion decay constant, if the scattering kernel is generalised to RML approximation, which is given by
\begin{eqnarray}\label{eq:piondecay}
f_{\pi_i} P_\mu = Z_2\int_{dq}{\rm tr}\left[i\gamma_5\gamma_\mu S(q_+)\Lambda_\beta\Gamma(q;P)\Lambda_\beta S(q_-)\right]\,,
 \end{eqnarray}
and the original pion decay constant definition with RL approximation, can be obtained from letting $\Lambda_\beta$ approach $\bf{1}$. A graphic representation of Eq.\eqref{eq:piondecay} is given in Fig.~\ref{fig:decayfeymann}. It is pointed out as well in Ref.~\cite{Naito:1998px} that if loop momentum cut-off or nonlocal interactions is included, pion decay constant must be modified accordingly, which is by analogy with our case herein. Practically, if considering RML in the case of $\Lambda^+_\beta$, owing to the existing of relations in Eq.\eqref{lambdaplustrace}, the Lorentz structure of pion decay constant on the right hand side of Eq.\eqref{eq:piondecay} is distinct from that with RL approximation. Additionally, solution of scalar function in the Bethe-Salpeter wave function,  contributing the most to the decay constant, $f_i(k;P)|i=2,3$ with $\Lambda^+_\beta$ is in general distinct from that with RL as well. Thus, combining these two effects, one cannot know in advance whether pion decay constant remains the same with RML in the case of $\Lambda^+_\beta$ compared to that with RL, and it requires further consideration from numerical computation. 

Given the normalised Bethe-Salpeter wave function, one can additionally consider the quantity associated with quark condensate with RML approximation, which is of the form
 \begin{eqnarray}\label{eq:pionrho}
 \rho_{\pi_i}=-Z_4\int_{dq}{\rm tr}\left[i\gamma_5 S(q_+)\Lambda_\beta\Gamma(q;P)\Lambda_\beta S(q_-)\right]\,, 
 \end{eqnarray}
and if considering RML in the case of $\Lambda^+_\beta$, it is noticed that owing to the existing of relations in Eq.\eqref{lambdaplustrace}, the Lorentz structure of $\rho_\pi$ on the right hand side of Eq.\eqref{eq:pionrho} is equivalent to that with RL approximation, so that one can expect before any numerical computation that $\rho_\pi$ with RML in the case of $\Lambda^+_\beta$ remains the same as that with RL. 

  In particular, the preservation of the axial-vector Ward-Green-Takahashi identity in Eq.\eqref{eq:avwti} yields the mass relation~\cite{Maris:1997hd}
\begin{eqnarray}\label{eq:gmor}
f_{\pi_i} M^2_{\pi_i}=2m^\zeta\rho_{\pi_i}(\zeta)\,,
\end{eqnarray}
which is known as the Gell-Mann-Oakes-Renner relation for ground state pion~\cite{GellMann:1968rz}. Additionally, this mass relation entails that leptonic decay constants of pion radial excitations vanish in chiral limit, and this is the consequence of chiral symmetry and its dynamical breaking in QCD~\cite{Holl:2004fr}. It is necessary for any development on the scattering kernel to preserve this relation, and any breaking of this relation might indicate the breaking of the underlying symmetry. We may hope our RML in the case of $\Lambda^+_\beta$ would preserve this relation, and the justification will be discussed in the following section.

\section{\label{sec:section5}Numerical results}

Our work so far has consisted of setting up a general quark-antiquark scattering kernel in the Bethe-Salpeter equation, the modified-ladder approximation, especially in the case of $\Lambda^+_\beta$, derived directly from the vector and axial-vector Ward-Green-Takahashi identities. One of the dominate applications of the modified-ladder approximation is the pion, and we have specified the procedure calculating pion mass, Bethe-Salpeter wave function, and their associated observables, i.e., decay constant, as well as the quantity associated with quark condensate. In other words, we have designated the scattering kernel, which is one of the pre-knowledge inputs required in constructing the Bethe-Salpeter equation. The remaining unknown input is the gluon propagator. Once scattering kernel and gluon propagator are both designated, the Bethe-Salpeter equation is well determined. Thus, in order to calculate pion properties, it now becomes necessary for us to implement the model of the gluon propagator. According to fruitful studies on this issue, we can apply the one introduced in Ref.~\cite{Qin:2011dd}, $D_{\mu\nu}(s)={\cal P}_{\mu\nu}{\cal G}(s)$: 
\begin{equation}\label{eq:gluonmodel}
 {\cal G}(s)=\frac{8\pi^2}{\omega^4}D e^{-s/\omega^2} +\frac{8\pi^2 \gamma_m \mathcal{F}(s)}{\text{ln}[\tau+(1+s/\Lambda^2_{QCD})^2]} \, ,
\end{equation}
where: ${\cal P}_{\mu\nu}=\delta_{\mu\nu}-\frac{p_{\mu}p_{\nu}}{p^2}$; $\gamma_m=12/(33-2N_{f})$, $N_{f}=4$, $\Lambda^{N_f=4}_{\text{QCD}}=0.234\,$GeV; $\tau=e^2-1$; and $\mathcal{F}(s)=[1-\exp(-s/[4m_t^2])]/s$, $m_t=0.5\,$GeV. The interaction in Eq.\eqref{eq:gluonmodel} involves a massive gluon scale on the domain at $s=0$, which is consistent with that determined in studies of QCD's gauge sector~\cite{Aguilar:2010gm,Binosi:2012sj,Binosi:2019ecz}. Parameters of interaction in Eq.\eqref{eq:gluonmodel} are taken as $D\omega=(0.82\,\rm GeV)^3$ and $\omega=0.5$ GeV, which is the typical  choice in a bulk of extant studies, and one can expect computed observables to be practically insensitive to the choice of $D$ or $\omega$ on a reasonable domain with keeping $D\omega$ stable~\cite{Chen:2018rwz}.

Additionally, practical calculation must take a renormalisation scale, at which physical quantities that we are interested are being considered. Of course physical observables such as pion mass and decay constant are independent of the chosen renormalisation scale. We take the renormalisation scale as $\zeta=0.3$ GeV herein, which is inspired by a recent progress on pion parton distribution function~\cite{Ding:2019lwe,Ding:2019qlr}, and the scale is originally from process independent running coupling~\cite{Binosi:2016nme,Rodriguez-Quintero:2018wma,Roberts:2020udq,Roberts:2020hiw}. RL requires the renormalisation-group-invariant light current-quark mass $\hat{m}=6.7$ MeV, which corresponds to $m^\zeta=12.7$ MeV. We take parameters with RML the same as those with RL, and in this way, all the distinctions of pion properties with RML in the case of $\Lambda^+_\beta$ and with RL can be considered as induced by the variation of scattering kernel.

\subsection{\label{sec:section5sub1}Masses of ground-state $\pi_0$ and first radial excited state $\pi_1$ }

After setting up computing inputs, let us consider our numerical results on the eigenvalue of the Bethe-Salpeter kernel, which corresponds to one at physical point when pion is on shell. Nambu-Goldstone theorem predicts that pion is massless in chiral limit. To verify whether our modified kernel is well-constructed, the first basic criteria we must consider is whether Nambu-Goldstone theorem is manifest with RML in the case of $\Lambda^+_\beta$. We have numerically verified that $\Lambda^+_\beta$ leads to a massless pion in chiral limit, the RML in the case of $\Lambda^+_\beta$ thus meets the first criteria, which is necessary to be manifest when developing a new scattering kernel.

\begin{figure}[t]
\centerline{\includegraphics[width=0.50\textwidth]{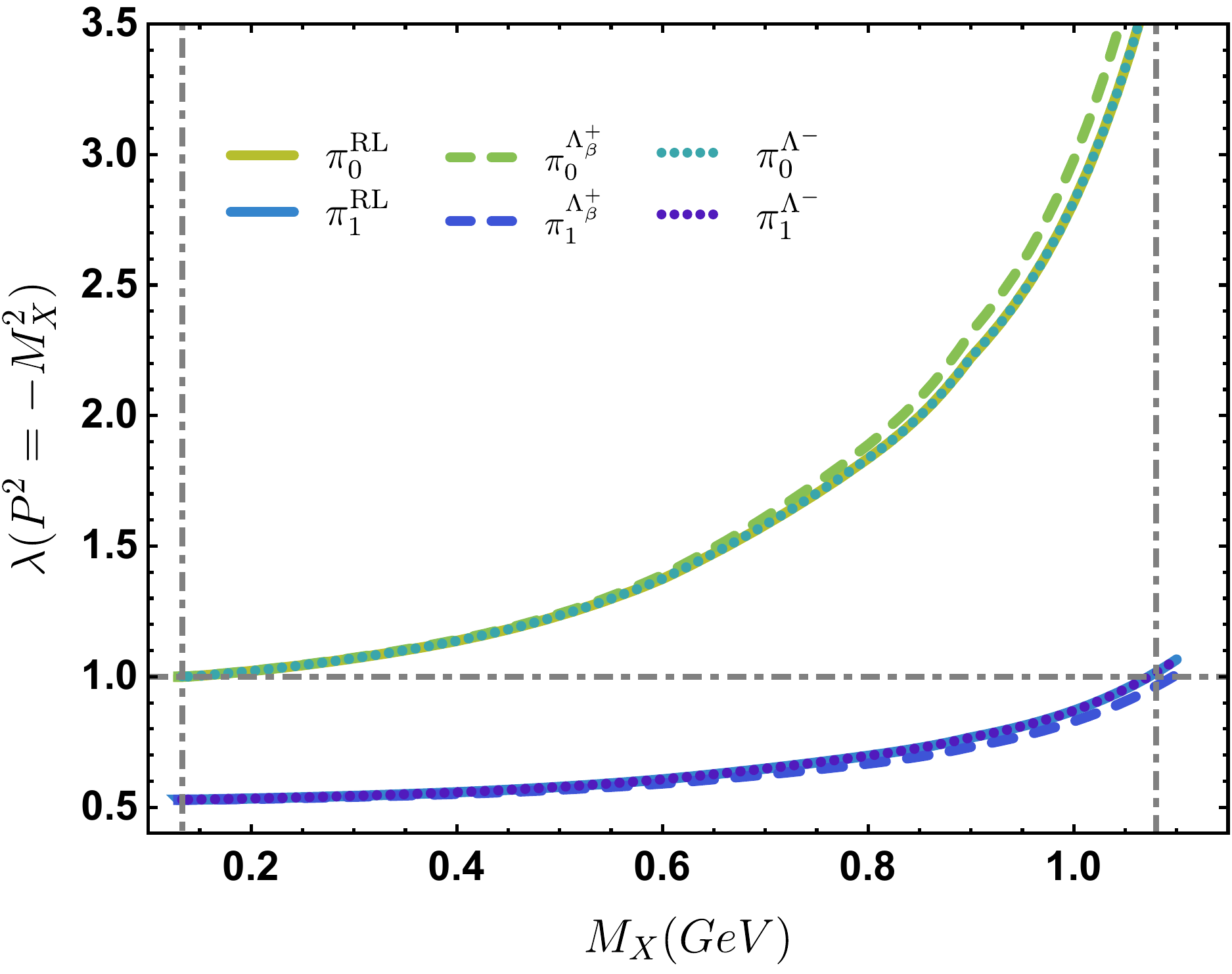}}
\caption{Eigenvalue dependences on meson mass of ground-state pion $\pi_0$, and first  radial excited state $\pi_1$  with RL and RML ($\Lambda^+_\beta,\,\Lambda^-$) approximations . $\emph{Solid}$ - RL; $\emph{Dashed}$ - RML in the case of $\Lambda^+_\beta$; $\emph{Dotted}$ - RML in the case of $\Lambda^-$. The  vertical lines indicate the location where the curves cross $\lambda=1$, corresponding to the physical masses of these two states.} \label{fig:eigenvalue}
\end{figure}

Further, let us consider pion mass beyond chiral limit. As we have seen in the previous section, eigenvalues of the Bethe-Salpeter kernel in homogeneous equations depend on meson mass, and physical states are those corresponding to $\lambda=1$. The eigenvalue dependences on meson mass of ground-state $\pi_0$, and first radial excited state $\pi_1$ with RL and RML are illustrated in Fig.~\ref{fig:eigenvalue}. We see that eigenvalue dependence pattern in meson mass is nearly identical among kernels. RML in the case of $\Lambda^-$ gives exactly equivalent eigenvalue dependence with RL, whereas $\Lambda^+_\beta$ show slightly distinction with them. This resembling pattern in eigenvalues of $\Lambda^-$ can be easily understood, as we have pointed out earlier that all equations in this case are equivalent to those with RL, therefore eigenvalues of the kernel will be the same as well. However the nearly resembling pattern in eigenvalues with $\Lambda^+_\beta$ is out of our expectation, since equations for $f_{2,3,4}(k;P)$ are completely different with $\Lambda^+_\beta$ and RL. Nevertheless, the eigenvalues behaviour suggests us that despite scattering kernels in various approximations affect the forms of Bethe-Salpeter equation, physical masses of the ground-state $\pi_0$ and first radial excited state $\pi_1$ are stable and nearly degenerate among distinct approximations
\begin{equation}
	M_{\pi_0}=0.133\, \text{GeV}\,,\quad\quad M_{\pi_1}=1.08\pm0.03 \,\text{GeV}\,.
\end{equation}
We include the variation for first radial excited state $\pi_1$ mass among kernels. Notably, the mass of $\pi_1$ herein is consistent with Ref.~\cite{Holl:2004fr}.

It is rather surprising to see nearly degenerate behaviour in both $\pi_0$ and $\pi_1$ masses among various approximations in the view of fact that their Bethe-Salpeter equations with RML in the case of $\Lambda^+_\beta$ and RL are in general not equivalent, there being apparent extra Dirac structures as a result of the inclusion of a multiplicative factor in the scattering kernel. The degenerate feature may be considered as the suggestion for a stable pion mass so long as the scattering kernel is constructed consistently. The qualification that has to be added here is  the requirement, that the kernel must be the solution of the symmetry-preserving vector and axial-vector Ward-Green-Takahashi identities. 

\subsection{\label{sec:section5sub2}Bethe-Salpeter amplitude of ground-state $\pi_0$}

\begin{figure}[t]
\centerline{\includegraphics[width=0.25\textwidth]{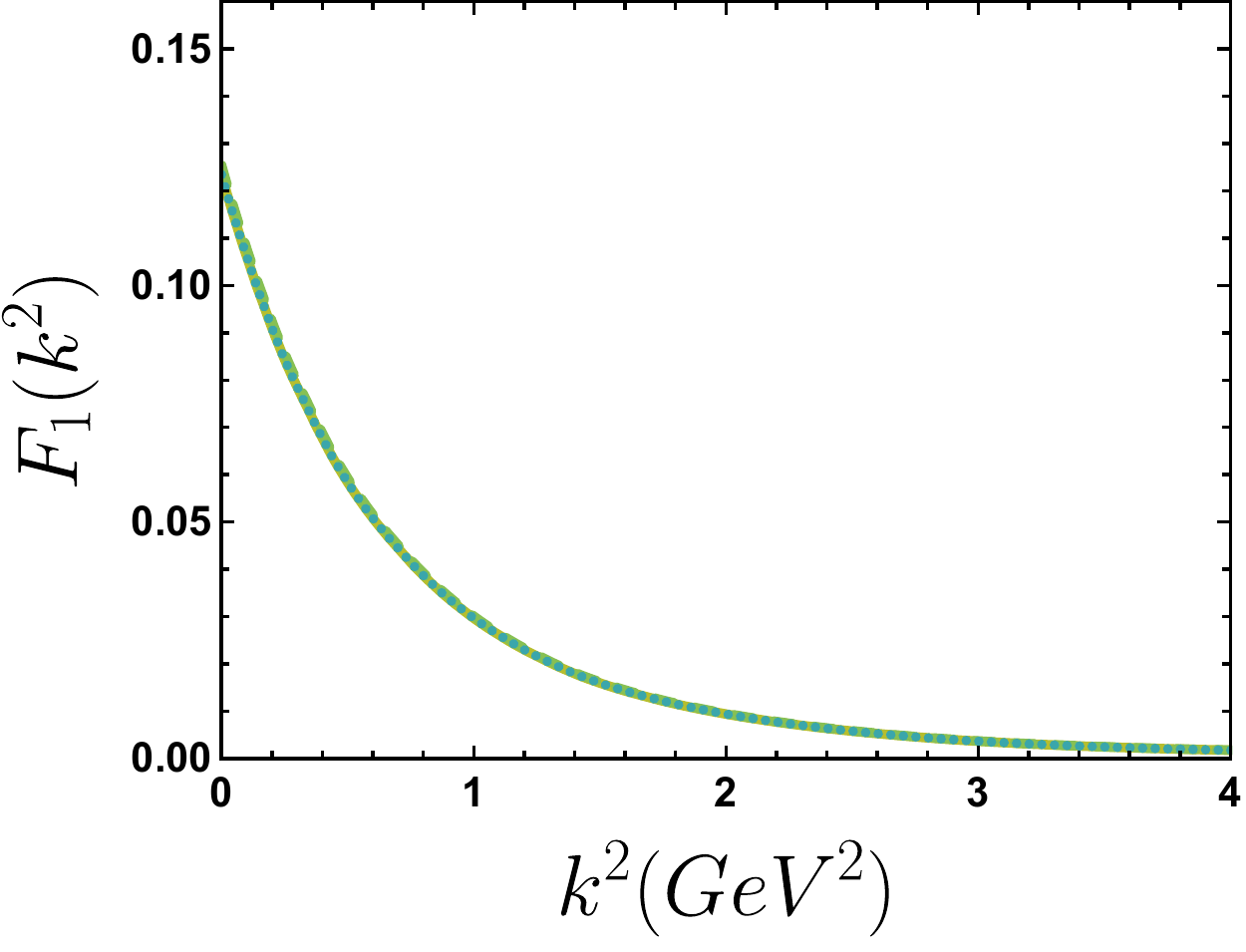}
\includegraphics[width=0.25\textwidth]{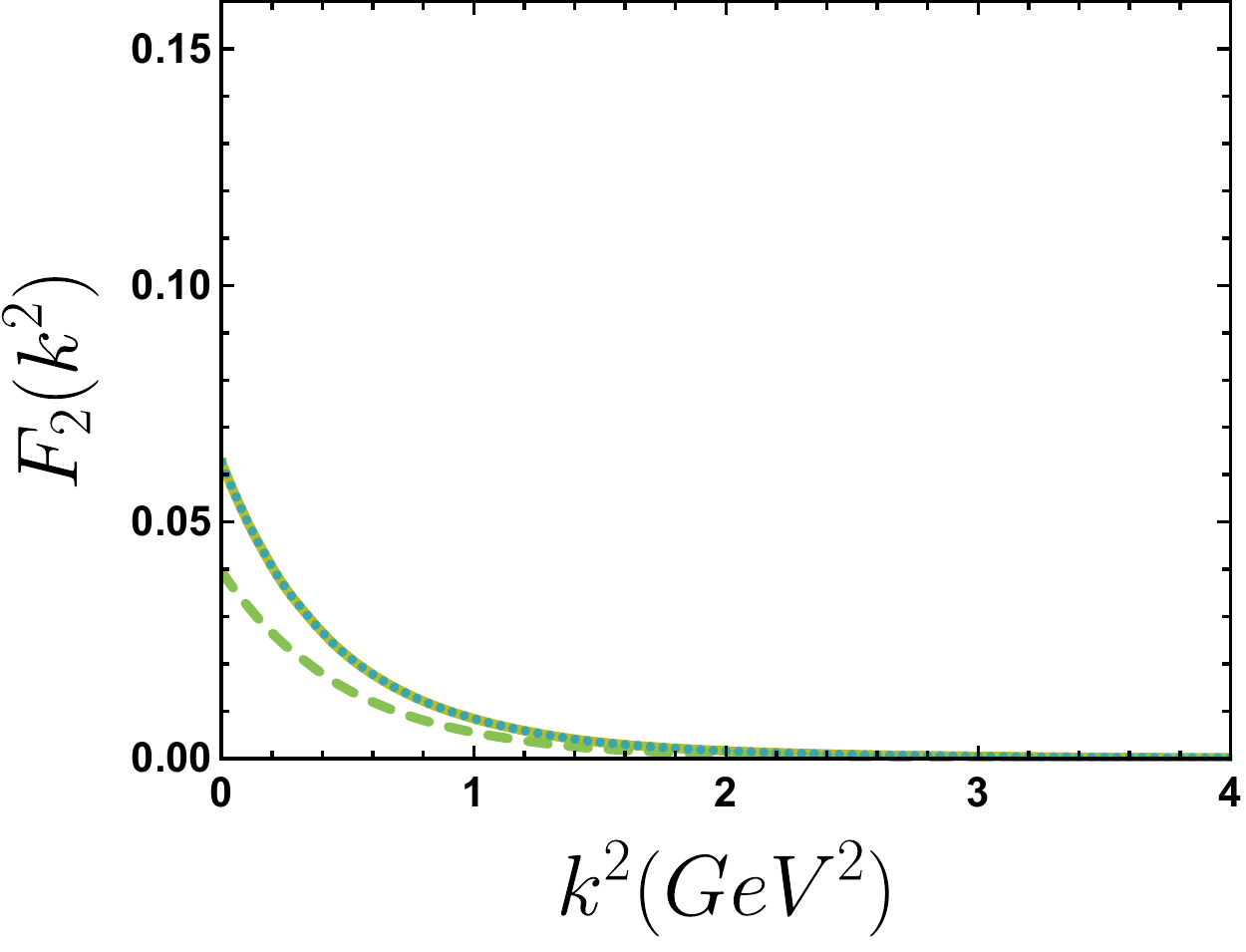}}
\centerline{\includegraphics[width=0.25\textwidth]{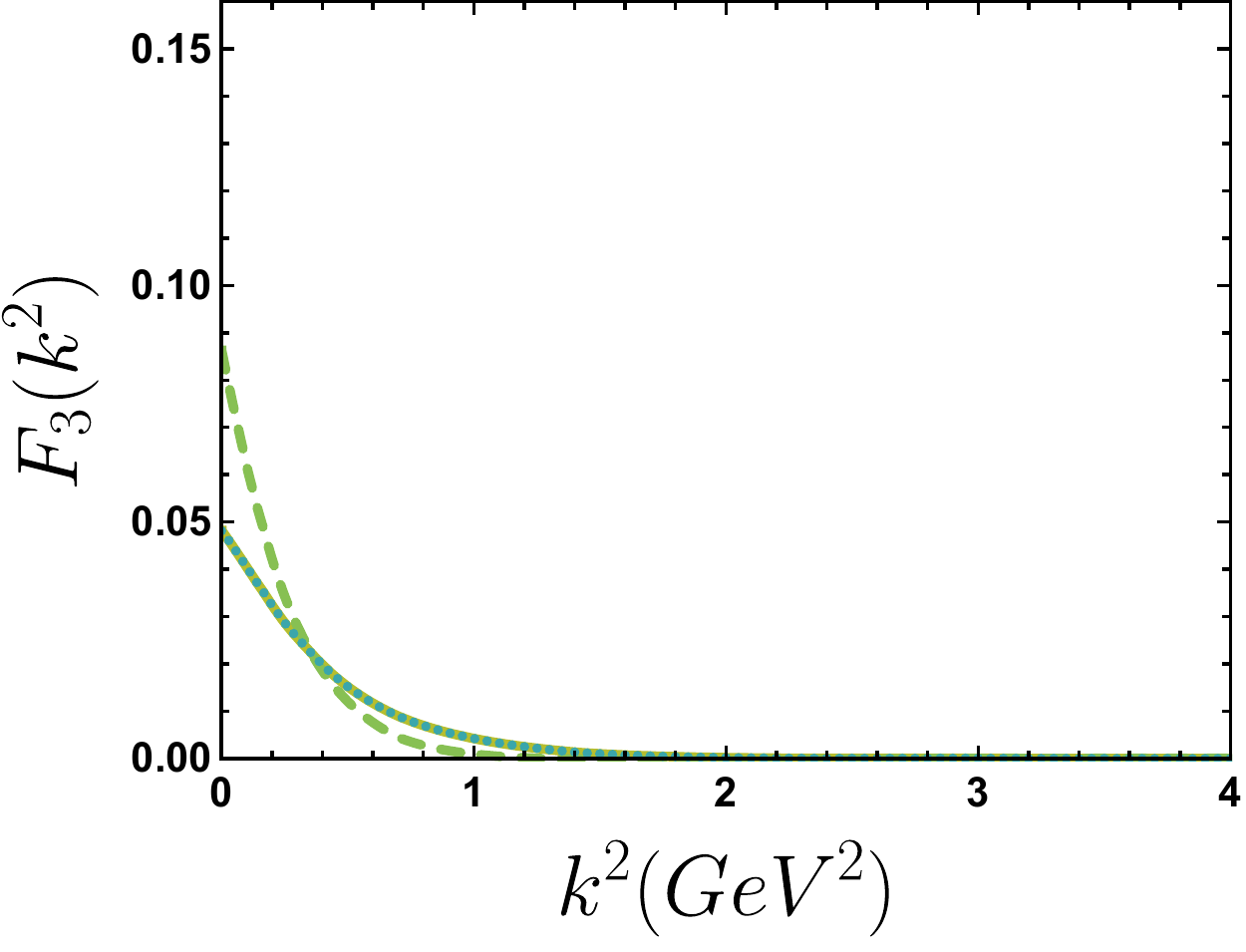}
\includegraphics[width=0.25\textwidth]{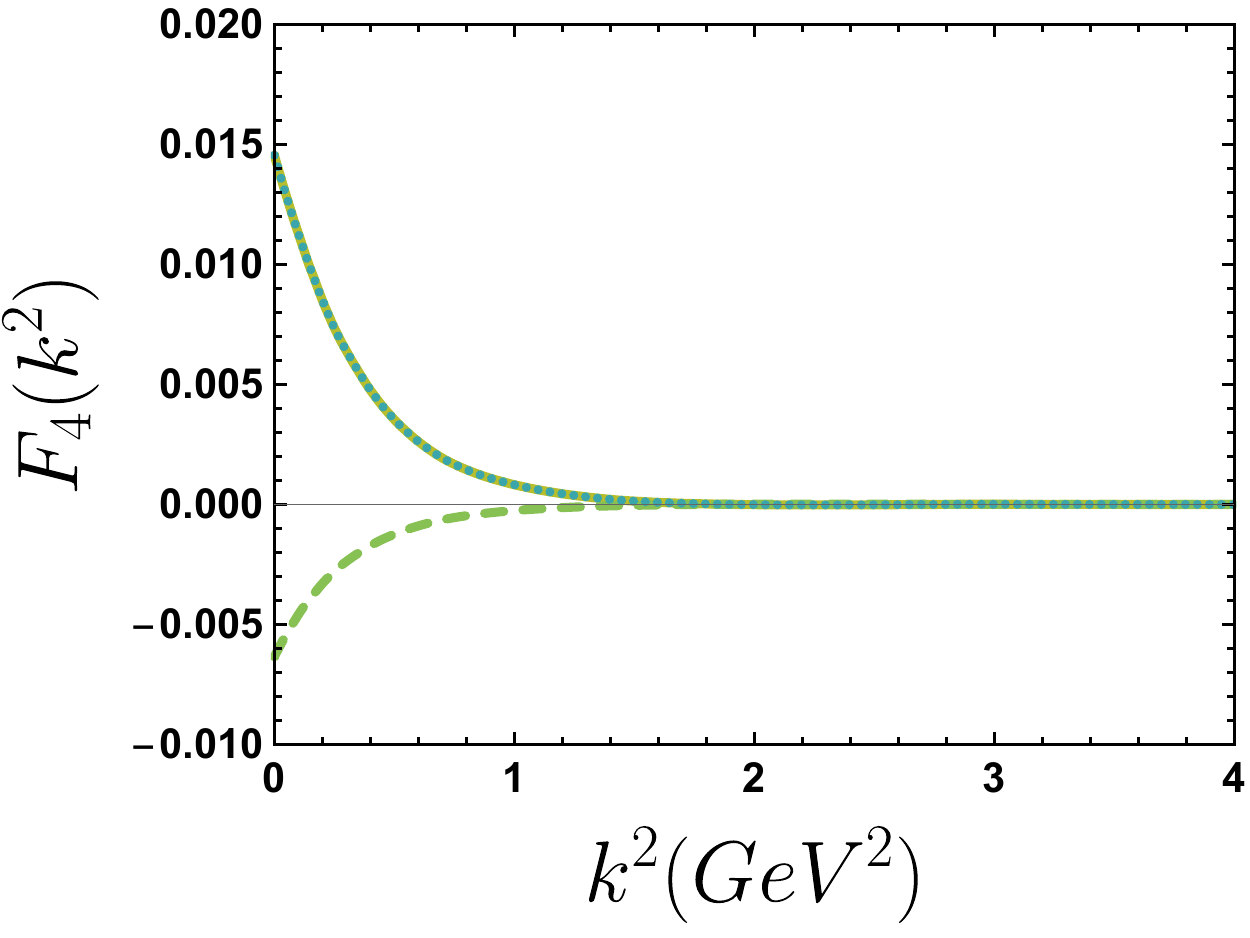}}
\caption{Lowest-order Chebyshev-projection of Bethe-Salpeter amplitudes $F_{1-4}(k^2)$ for ground-state pion $\pi_0$. In all panels, $\emph{Solid}$ - RL; $\emph{Dashed}$ - RML in the case of $\Lambda^+_\beta$; $\emph{Dotted}$ - RML in the case of $\Lambda^-$. $\emph{Row-1}$, $\emph{left}$, $F_1$; $\emph{Row-1}$, $\emph{right}$, $F_2$; $\emph{Row-2}$, $\emph{left}$, $F_3$; and $\emph{Row-2}$, $\emph{right}$, $F_4$.} \label{fig:BSA0}
\end{figure}

Up to this point we have seen results for eigenvalues of the Bethe-Salpeter kernel, which are associated with pion masses at physical point, we now continue to consider eigenvectors, $i.e.$, detailed structure of the Bethe-Salpeter amplitude. As an immediate example, we look attentively at the ground state $\pi_0$. Pion non-perturbative properties are all carried by its Bethe-Salpeter amplitude, so that the amplitude is fundamental itself for those attempting to describe pion internal structure~\cite{Maris:1997tm,Qin:2011xq}. If we write pion Bethe-Salpeter amplitude as $\Gamma(k;P)=\sum_{i=1}^4 \tau_i(k;P) F_i(k;P)$, with $\tau_i(k;P)$ is given in Eq.\eqref{eq:pionbsebasis}, then $F_i(k,P)$ is the scalar function characterising pion amplitude dependence on relative and total momentum of dressed quark and dressed antiquark. Numerical results for the lowest-order Chebyshev-projection of $F_i(k,P)$  
\begin{align}
F_i(k^2):=\frac{2}{\pi}\int^1_{-1}dx\sqrt{1-x^2}U_0(x)F_i(k^2,x;P^2)\,,	
\end{align}
for ground-state $\pi_0$ with both RL and RML ($\Lambda^+_\beta$, $\Lambda^-$) are shown in Fig.~\ref{fig:BSA0}, where $k\cdot P=x\sqrt{k^2P^2}$ and $U_0(x)$ is the lowest-order Chebyshev polynomial of the second kind.

One apparent feature in all panels of Fig.~\ref{fig:BSA0} is the resembling pattern among all the lowest-order Chebyshev-projection of Bethe-Salpeter amplitudes $F_{1-4}(k^2)$ with RML in the case of $\Lambda^-$ (dotted curve) and those associated with the conventional RL (solid curve) approximation. According to the proof described above, all the Bethe-Salpeter equations are equivalent with these two cases, for this reason, we should obtain equivalent eigenvectors of the Bethe-Salpeter kernel. Numerical results herein have verified this statement. Therefore we will focus ourselves on reviewing the difference on numerical results of Bethe-Salpeter amplitude between RML in the case of $\Lambda^+_\beta$ and RL.

Unlike the nearly degenerate feature in pion masses as we described above, the lowest-order Chebyshev-projection of Bethe-Salpeter amplitude of $\pi_0$ expresses distinct behaviour with RML in the case of $\Lambda^+_\beta$ (dashed curve) and RL (solid curve). In detail, $F_1$ turns out to be equivalent in both cases, whereas $F_{2,3,4}$ behaves differently with their dependence on relative momentum of the dressed quark and dressed antiquark in the infrared region. If we consider RML in the case of $\Lambda^+_\beta$, the general decreasing behaviour of $F_2$ remains, whereas $F_2(k^2=0)$ is relatively smaller in comparison to that with RL. $F_3$ keeps the deceasing behaviour as well, and $F_3(0)$ is relatively larger. $F_4$ turns out to be negative. 

One might find explanations for these features from equations we have mentioned above. Equations for $F_1$ with RML in the case of $\Lambda^+_\beta$ is equivalent to those with RL, whereas equations for $F_{2,3,4}$ with $\Lambda^+_\beta$ differ from the ones with RL (see Eq.\eqref{lambdaplustrace}). Consequently, it may be possible that these two approximations yield the same $F_1$ yet different $F_{2,3,4}$. The appearance of different $F_{2,3,4}$ can be employed to analyse the content of the Bethe-Salpeter wave function with a specific orbital angular momentum~\cite{Hilger:2015ora}. These features in the Bethe-Salpeter amplitude indicate that in these two approximations it is possible to have various s- and p-wave contributions, meson structure functions will have more general properties rather than the unique behaviour.

\subsection{\label{sec:section5sub3}Decay constant of ground-state $\pi_0$ and the GMOR relation}

The distinction in ground-state $\pi_0$ Bethe-Salpeter amplitude with RML in the case of $\Lambda^+_\beta$  and RL may lead to distinct decay constant as defined in Eq.\eqref{eq:piondecay}. However, in practice we find
\begin{align}
	f^{\text{RL}}_{\pi_0}=92.3\, \text{MeV}\,,\quad\quad f^{\Lambda^+_\beta}_{\pi_0}=93.5 \,\text{MeV}\,,
\end{align}
in comparison to that in experiment $f^{\text{Exp}}_{\pi_0}=91.9\pm3.54$ MeV~\cite{Zyla:2020zbs}. Therefore one may notice that both $f^{\text{RL}}_{\pi_0}$ and $f^{\Lambda^+_\beta}_{\pi_0}$ are compatible to the experimental value by relative errors within $2\%$. Again it is rather surprising to see nearly degenerate pion decay constant among various approximations in the view of fact that their Bethe-Salpeter amplitude are in general not equivalent (see Fig.~\ref{fig:BSA0}). The degenerate feature may also be considered as the suggestion for a stable pion decay constant so long as the scattering kernel is constructed from Ward identities.  Pion decay constant, together with pion mass is physical observable, which should be in principle invariant among models, and our practical calculation shows this is the case herein. Other physical observable, such as pion parton distribution function, we can expect it will also be invariant with any symmetry-preserving modification of the Bethe-Salpeter scattering kernel, and this remains justification by future studies.

In order to examine whether the perseveration of GMOR relation still holds with RML in the case of $\Lambda^+_\beta$ and RL approximations, we include the calculation the quantity associated with quark condensate defined in Eq.\eqref{eq:pionrho}, and obtain
\begin{align}
	\rho^{\text{RL}}_{\pi_0}(\zeta)=(0.252\, \text{GeV})^2\,,\,\, \rho^{\Lambda^+_\beta}_{\pi_0}(\zeta)=(0.250 \,\text{GeV})^2\,.
\end{align}
We have seen in the preceding section that $\rho_{\pi_0}$ with $\Lambda^+_\beta$ is equivalent to that with RL based on analytical analysis, and practical numerical results turn out to be consistent with this statement. Now we can consider the maintenance of the GMOR relation in Eq.\eqref{eq:gmor}. If it is maintained, then the ratio of decay constant $f_{\pi_0}$ and $\rho_{\pi_0}$ is
\begin{align}\label{eq:gmorratio}
	{f_{\pi_0}}/{\rho_{\pi_0}}(\zeta)={2m^\zeta}/{M^2_{\pi_0}}\,.
\end{align} 
In practice, we use the same light current quark mass $m^\zeta=12.7$ MeV with RML in the case of $\Lambda^+_\beta$ and with RL, and we additionally notice that ground-state $\pi_0$ mass degenerate $M_{\pi_0}=0.133$ GeV in these two approximations, therefore the ratio on the right hand side of Eq.\eqref{eq:gmorratio} with $\Lambda^+_\beta$ is exactly equivalent to that with RL, $viz.$, ${2m^\zeta}/{M^2_{\pi_0}}=1.436$ GeV$^{-1}$. When considering the left hand side of Eq.\eqref{eq:gmorratio} with these two approximations, we find
${f^{\text{RL}}_{\pi_0}}/{\rho^{\text{RL}}_{\pi_0}}(\zeta)=1.451 \,\text{GeV}^{-1}$, ${f^{\Lambda^+_\beta}_{\pi_0}}/{\rho^{\Lambda^+_\beta}_{\pi_0}}(\zeta)=1.496\, \text{GeV}^{-1}$, hence the GMOR relation is preserved with RL and RML by a relative error within $5\%$.

Notably, considering RML in the case of $\Lambda^+_\beta$, the GMOR relation is preserved only if we follow the definition of pion decay constant in Eq.\eqref{eq:piondecay} with a graphic representation  given in Fig.~\ref{fig:decayfeymann}, and the definition of $\rho_{\pi_0}$ in Eq.\eqref{eq:pionrho}, in which the multiplicative factor is included. This might suggest us that when considering physical observables making use of the Bethe-Salpeter amplitude with the modified-ladder approximation, one must take into account the possible modification on the expression associated with the interested physical observables. 

Additionally, given the GMOR relation connects current quark mass, pion mass, decay constant and the quantity associated with quark condensate, one can start from it and obtain above quantities from one to anther. However, if a theory can give these  quantities simultaneously, it would be not easy for them to satisfy the GMOR relation with a high accuracy. Therefore, besides the Nambu-Goldstone theorem, the GMOR relation can serve as a second criteria, which is necessary to manifest for pion when developing a new scattering kernel.

\section{\label{sec:section6}Conclusion}
In this paper, we explored the possible rainbow modified-ladder approximation, derived directly from the vector and axial vector Ward-Green-Takahashi identities. Starting from Ward identities in rainbow approximation, we obtain two equations for the quark-antiquark scattering kernel in the Bethe-Salpeter equation, Eq.\eqref{eq:relationfromvwti} and Eq.\eqref{eq:relationfromavwti}. Then the quark-antiquark scattering kernel is assumed to include a multiplicative factor in comparison to that with the conventional rainbow ladder approximation. Apart from $\bf{1}$, two nontrivial solutions for the multiplicative factor are found, as outlined in Eq.\eqref{eq:models}, corresponding to the rainbow modified-ladder approximation. There is a distinction with the modified-ladder approximation scattering kernel in comparison to that in ladder approximation, it owns momentum dependence described by the quark momentum and/or the vector part of quark propagator. In the consequence of this, it may lead to some impact on properties of the system one is interested. 

As an application of this rainbow modified-ladder approximation, we study pion. We have first numerically verified that the Nambu-Goldstone theorem is manifest with the modified-ladder approximation. Then the most remarkable result is that the pion masses of both ground-state $\pi_0$ and first radial excited state $\pi_1$ are degenerate with rainbow ladder and rainbow modified-ladder approximations, as illustrated in Fig.~\ref{fig:eigenvalue}. The degenerate feature may suggest for a stable pion mass so long as the scattering kernel is constructed from the symmetry-preserving Ward identities. The Bethe-Salpeter amplitude of the ground-state $\pi_0$ is then considered. Unlike the degenerate feature in pion masses, the Bethe-Salpeter amplitude of $\pi_0$ expresses distinct behaviour with RML in the case of $\Lambda^+_\beta$ and RL, as given in Fig.~\ref{fig:BSA0}. It is then noticed that the distinction on $\pi_0$ Bethe-Salpeter amplitude does not affect pion decay constant, in the case RML, it is generally equivalent to that with RL. Consequently, the GMOR relation is also numerically verified to be preserved with the modified-ladder approximation.

The complexity of numerical computation with rainbow modified-ladder approximation does not increase dramatically compared to rainbow ladder approximation, so that the study herein can be easily extended to further investigation on other mesons such as heavy quarkonium and mesons with nonzero spin.

\begin{acknowledgments}
We acknowledge valuable input from Daniele Binosi, Muyang Chen, Fei Gao, and Craig Roberts.
 
\end{acknowledgments}

\nocite{*}

\bibliography{RML.bib}

\end{document}